\documentclass[runningheads]{llncs}
\usepackage{hyperref}
\usepackage{amsfonts}
\usepackage{amsmath}
\usepackage{booktabs}
\usepackage{multirow}

\usepackage{graphicx}
\usepackage{tikz, pgfplots, pgfplotstable, placeins, subcaption, rotating}
\pgfplotsset{compat=1.14}

\begin{document}
\title{Blockchain Price \textit{vs.} Quantity Controls}
 \author{Abdoulaye Ndiaye\orcidID{0009-0000-7466-6444}}
 \authorrunning{A. Ndiaye}
 \institute{New York University, New York NY 10012, USA\\
 \email{andiaye@stern.nyu.edu}\\
 \url{https://www.abdoulayendiaye.com/}}

\maketitle
\begin{abstract}
This paper studies the optimal transaction fee mechanisms for blockchains, focusing on the distinction between price-based ($\mathcal{P}$) and quantity-based ($\mathcal{Q}$) controls. By analyzing factors such as demand uncertainty, validator costs, cryptocurrency price fluctuations, price elasticity of demand, and levels of decentralization, we establish criteria that determine the selection of transaction fee mechanisms. We present a model framed around a Nash bargaining game, exploring how blockchain designers and validators negotiate fee structures to balance network welfare with profitability. Our findings suggest that the choice between $\mathcal{P}$ and $\mathcal{Q}$ mechanisms depends critically on the blockchain’s specific technical and economic features. The study concludes that no single mechanism suits all contexts and highlights the potential for hybrid approaches that adaptively combine features of both $\mathcal{P}$ and $\mathcal{Q}$ to meet varying demands and market conditions.
\end{abstract}


\section{Introduction} 

Transaction fee mechanisms play a crucial role in maintaining blockchains' network stability, efficiency, and user satisfaction. These policies can be broadly categorized into three main types, each with distinct characteristics and implications for the network's economic and operational dynamics.

Quantity Controls ($\mathcal{Q}$): This approach involves setting a maximum limit on resource usage, such as the block size limit. Bitcoin's fee policies are a prime example of this strategy, where the block size limit is the primary mechanism controlling the volume of transactions processed in each block. By capping the block size, the network effectively manages transaction throughput through a first-price auction.

Price Controls ($\mathcal{P}$): Under this mechanism, a minimum price per unit of resource usage is set, often dynamically adjusted to reflect current network conditions. Ethereum's EIP-1559 and  EIP-4844 illustrate cases where a base fee is determined algorithmically, rising or falling based on the network's congestion levels. This approach allows the quantity of transactions to adjust in response to demand changes.

Fixed Unit Price ($\emptyset$): Some networks implement a fixed fee structure without adjusting for market conditions. This straightforward approach, often adopted at the beginning of the life-cycle of blockchains, can simplify transactions for users but may lack the flexibility needed to address fluctuating network demand effectively.

The selection between these transaction fee mechanisms—whether a blockchain opts for quantity control ($\mathcal{Q}$), price control ($\mathcal{P}$), or a fixed price approach ($\emptyset$)—is influenced by several factors. Key among these is the interplay between the objectives of welfare-maximizing blockchain designers and profit-maximizing validators. Blockchain designers typically aim to enhance overall network efficiency and user satisfaction, while validators are incentivized to maximize their earnings from transaction fees and block rewards.

The central question in this paper is: What specific conditions or characteristics of a blockchain affect the decision between implementing a price control or a quantity control mechanism? By analyzing various blockchain architectures and economic factors, such as uncertainty in user demand, uncertainty in costs to a validator to process transactions, cryptocurrency price fluctuations, 
price elasticity of demand, and levels of decentralization, we establish criteria that determine the selection of transaction fee mechanisms.

Table \ref{tab:summary} summarizes those criteria. First, in environments with high demand uncertainty, exemplified by blockchains with various use cases, following a price control is preferred to adjust the block size to match fluctuations in demand. Second, a quantity control is favored for blockchains with a consensus mechanism, such as Proof of Work (PoW), where there is a significant positive correlation between demand uncertainty and marginal costs (hash rate). The reason is that if the block size limit was allowed to adjust, validators may face higher costs when demand—and hence their workload—increases. Third, when token prices fluctuate widely, implementing quantity controls helps avoid base fees that are too high and leads to more stable transaction fees denominated in the native token. Fourth, blockchains characterized by a high elasticity of demand for block space, such as those with faster blocks or quicker confirmations, benefit from price controls, which allow more flexible and responsive fee adjustments. Fifth, quantity controls are adequate in highly decentralized networks with low validator bargaining power, as they become easier to enforce.

The optimal choice between an EIP-1559 type policy ($\mathcal{P}$) or a traditional block size limit ($\mathcal{Q}$) is determined by the relative balance of these five economic factors that emerge from the blockchain technical specificities. Both Bitcoin and Ethereum blockchains face significant uncertainties in user demand. The marginal cost for Ethereum validators has been essentially constant since the Proof of Stake (PoS) upgrade. In addition, Ethereum features faster blocks and is arguably less decentralized than Bitcoin. Therefore, our results help explain Ethereum's adoption and planned adoption of price control mechanisms such as EIP-1159 and EIP-4844. At the same time, they help explain why Bitcoin still uses a sole block size limit as quantity control.
\begin{table}[h!]
    \centering
    \renewcommand{\arraystretch}{1.5} %
    \begin{tabular}{c|c|c}
        \textbf{FACTORS} & \textbf{EXAMPLE} & \textbf{OUTCOME}\\
        \hline \hline
        
        High demand uncertainty & Different use cases & $\mathcal{P}$\\
        + Corr. btw demand uncertainty and MC & PoW & $\mathcal{Q}$\\
        
        Token price fluctuations & Fees  in native token & $\mathcal{Q}$ \\
                    High elasticity of inclusion in next block & Faster blocks or confirmation & $\mathcal{P}$\\
        Low validator bargaining power & High decentralization & $\mathcal{Q}$\\
        \hline

    \end{tabular}
    \medskip
    \caption{Summary of factors leading to a $\mathcal{P}$ or $\mathcal{Q}$ equilibrium between "welfare-maximizing" blockchain designers and "profit-maximizing" validators.}
    \label{tab:summary}
\end{table}
\subsection{Literature Review}
This research builds upon the foundational studies on price versus quantity controls initiated by ~\cite{weitzman1974prices}  The choice of selecting a supply function under uncertain conditions has been extensively discussed in the work of ~\cite{klemperer1989supply}  My methodology closely resembles the analysis by ~\cite{reis2006inattentive}, who investigates the implications of these choices from a macroeconomic standpoint  However, my approach diverges by focusing on the unique challenges blockchain designers face, who must balance multiple technical and strategic objectives, such as setting block size limit The results of this study contribute to the development of transaction fee mechanisms.

The microeconomic mechanism design perspective on TFMs has seen substantial growth, particularly with the contributions of ~\cite{akbarpour2020credible}, who analyze "credible mechanisms" that resist manipulation by designers. These are particularly relevant in the blockchain sphere, where ~\cite{roughgarden2021transaction} demonstrates that Ethereum's EIP-1559 and related models meet these criteria, offering a myopically credible solution for validators and users. This is further supported by the findings of ~\cite{chung2023foundations}, ensuring that the TFMs underlying the price controls studied here are incentive-compatible.

From a broader macroeconomic angle, this paper applies results in ~\cite{ndiaye2023bitcoin} to enhance our understanding of the tradeoffs in choosing the families of blockchain TFMs.

\paragraph{Outline:} The paper is organized as follows: Section \ref{sec: Model} provides intuition on the economics of price and quantity controls with perfect enforcement. Section \ref{sec: results} presents how each factor affects the choice of controls in the blockchain context where protocol designers cannot fully enforce policies.
Section \ref{sec: conclusion} takes stock of the results, discusses the general choice of a supply function for block space, and concludes the paper.
 \section{Prices vs. Quantity Controls with Perfect Enforcement} \label{sec: Model}

In this section, we analyze the welfare implications of price and quantity controls amid demand uncertainty and social cost uncertainty in a general setting to illustrate the idea of ~\cite{weitzman1974prices}.

Consider a general private benefit of block space $q$, denoted by $B(q)$. These private benefits may reflect the utility users derive from generating a block of size $q$ that matches their total demand. Social costs could include costs beyond those incurred by validators, such as centralization costs. Uncertainty in private benefits and social costs are introduced through the $B(q,\Psi)$ and $C(q,\eta)$. From the perspective of the blockchain designer, the value of setting the quantity  $\bar{q}$ in advance is given by:
\begin{equation}
\max_{\bar{q} \in \mathbb{R}_{+}}\mathbb{E}_{\Psi,\eta}[B(\bar{q},\Psi)-C(\bar{q},\eta)].
\end{equation}
Alternatively, if the blockchain designer sets the price $\underline{p}$, in advance, quantities adjust ex-post to match demand:
\begin{equation}
B_1(q^{adj}(\underline{p},\Psi),\Psi)=\underline{p}.
\end{equation} 
The value of this price control takes quantity adjustments into account:
\begin{equation}
\max_{\underline{p} \in \mathbb{R}_{+}}\mathbb{E}_{\Psi,\eta}[B(q^{adj}(\underline{p},\Psi),\Psi)-C(q^{adj}(\underline{p},\Psi),\eta)],
\end{equation}

For intuition on the choice of instruments between a minimum price (base fee) and a maximum quantity (block size limit), consider a setting where there is no uncertainty in the cost function but with a 50\% chance demand is high and a 50\% chance demand is low  Figure \ref{fig:price_control} illustrates this example with a quantity limit (top panel) and price control (bottom panel) under such demand uncertainty.
\begin{figure}[htbp]
\centering
\input{./figures/dwl_q.tex}
\end{figure}
\begin{figure}
\centering
\input{./figures/dwl_p.tex}
\caption{Welfare improvement from price controls under demand uncertainty. Top panel: deadweight loss of a quantity limit. Bottom panel: deadweight loss of price control.}
\label{fig:price_control}
\end{figure}

Consider, without loss of generality, that the block size limit $q^{\max}$ is equal to block space demand when demand is low but is binding when demand is high, as depicted in the top panel of the figure. In this context, the concept of deadweight loss comes into play. Deadweight loss refers to the loss in total social surplus due to an inefficient market outcome—it occurs when supply and demand are not in equilibrium. Here, the deadweight loss is represented by the shaded area between the high demand curve, the marginal cost curve, and the block size limit. This area is twice the deadweight loss that results from the block size limit.

Now, consider a scenario where the blockchain designer introduces a minimum price that exceeds the low-demand market price, as shown in the bottom panel of the figure. In this scenario, the dark-shaded area represents twice the deadweight loss that results from this price control. It can be observed that the deadweight loss from the price control is lower than the deadweight loss from the block size limit. This situation arises whenever the uncertainty in demand exceeds the uncertainty in marginal costs.

Formally, the demand curve can be approximated around a quantity limit $\bar{q}$ as:
\begin{equation}
B_1(q,\Psi) \approx B' + \beta(\Psi) + B'' \cdot (q - \bar{q}),
\end{equation}
and the marginal cost is approximated by:
\begin{equation}
C_1(q,\eta) \approx C' + \eta(\eta) + C'' \cdot (q - \bar{q}),
\end{equation}
where it is assumed that there is uncertainty around a fixed demand curve $B_1(q) \approx B' + B'' \cdot (q - \bar{q})$ and marginal cost curve $C_1(q) \approx C'  + C'' \cdot (q - \bar{q})$ such that $\mathbb{E}[\beta] = \mathbb{E}[\eta] = 0$.

The comparative advantage of a price control over a quantity control, denoted by $\Delta$, can be expressed as:
\begin{equation}
\Delta \equiv \mathbb{E}[(B(\tilde{q}(\Psi),\Psi) - C(\tilde{q}(\Psi),\eta)) - (B(\bar{q},\Psi) - C(\bar{q},\eta))].
\end{equation}

This comparative advantage of price control is:
    \begin{equation}
\Delta \propto \frac{B''}{C''^2} + \frac{1}{C''},
\end{equation}
and if $|B''| > C''$, a price control improves welfare over a quantity control.

This is the main result of ~\cite{weitzman1974prices}. When demand is more uncertain than marginal cost, price controls can lead to quantity adjustments that better match demand, while marginal costs do not vary much. This result determines when price controls lead to welfare improvements over quantity controls.

\subsection{EIP-1559: The Ethereum $\mathcal{P}$ Transaction Fee Mechanism}

Inspired by Weitzman's work about environmental regulation, ~\cite{buterin2018blockchain} introduced a revised pricing mechanism, EIP-1559, for the Ethereum blockchain. The system has a target block size, currently set at $q^{target} = 15M\text{ gas}$ (the unit of block size), and a maximum block size of $q^{\max} = 2 q^{target}$  The minimum gas price, $p_t$, is adjusted based on the formula
\begin{equation}\label{eq: EIP1559}
    p_t = p_{t-1} \cdot (1 + d \frac{q_{t-1} - q^{target}}{q^{target}})
\end{equation} where the adjustment parameter $d$ is set by default for the minimum price to double in 8 blocks when blocks are full, i.e., $d=\frac{1}{8}$. 

In the Ethereum blockchain, each transaction indexed by $j$ has an associated gas limit $q_j$, computed based on a fixed fee schedule $p_x$  Transaction senders pay an amount in ETH, the native currency of the blockchain, equal to $q_j \cdot \min\{p_t + \delta_j, c\}$, where $\delta_j$ is the tip and $c$ is the fee cap, with $c \geq p_t$  The base fee revenue, $\sum_{j=1}^N q_j p_t$, s burned, mainly for reasons related to off-chain incentives of validators ~\cite{roughgarden2021transaction}, while the effective tips are transferred to the validator.

\subsection{Modeling Key Differences between Blockchains Fees and Environment Regulation}

Blockchains, particularly in permissionless systems, present unique challenges that differentiate them from traditional economic regulation under uncertainty, as illustrated above. These challenges stem primarily from the decentralized nature of blockchains and the diverse stakeholders involved, each with different objectives and influences on market equilibrium.

    \paragraph{Diverse Stakeholders:} The key actors in a blockchain ecosystem include developers, validators (proposers or builders), and users. Each group holds varying degrees of power and influence over the network's operations and policies.
    \paragraph{Absence of Central Authority:} Unlike a government, a blockchain designer cannot unilaterally enforce policies. Validators, who play a critical role in processing transactions and creating new blocks, must be incentivized to follow proposed changes, which may not always align with their interests.
    \paragraph{Uncertainty:} Blockchains face multiple sources of uncertainty that affect their operation and the feasibility of different transaction fee mechanisms. These include fluctuations in user demand, the variable costs faced by validators, and the volatile prices of cryptocurrencies.

To address these challenges, we propose in ~\cite{ndiaye2023bitcoin} a model that conceptualizes the decision-making process regarding transaction fee mechanisms as a Nash bargaining game between blockchain designers and validators. The model is structured around the following components:

    \paragraph{Decisions:} Blockchain designers must commit ex-ante to either a fixed base fee ($\mathcal{P}$-setting) or a fixed blockspace ($\mathcal{Q}$-setting) before the full extent of uncertainties is realized.
    \paragraph{Bargaining Model of Decentralization:} The bargaining game is formalized by the following optimization problem:
    \begin{align}
    \max_{\mathcal{P},\mathcal{Q}}\mathbb{E}[\textrm{Social Benefit}(\Psi,\eta)]^{1-\beta}\mathbb{E}[\textrm{Validator Profits}(\Psi,\eta)]^{\beta}
    \end{align}
    where $\beta \in [0,1]$ represents the bargaining power of validators.

\section{How Different Factors Affects the $\mathcal{P}$ vs. $\mathcal{Q}$ choice}\label{sec: results}
In this section, we discuss the contribution of each factor in the choice of price or quantity controls.
\subsection{ Uncertainty in User Demand}

\subsubsection{Definition and Examples:}

In our model, user demand for blockchain transactions can be defined by the equation:
\begin{align}
 q = \frac{\Psi}{p^{\varepsilon}}   
\end{align}

Where $q$ represents the resource quantity used by transactions, $\Psi$ captures factors influencing demand such as transaction utility or network activity, and $p$ is the price per transaction. 

Demand uncertainty is quantitatively expressed through the variance of $\Psi$, denoted as $Var[\Psi]$. High variability in $\Psi$ indicates high uncertainty in user demand. This variability can be attributed to several causes:
    \paragraph{Different Use Cases:} Blockchains can serve various applications with distinct demand patterns.
    \paragraph{Adoption Phase:} As the technology matures and gains wider acceptance, demand for block space can increase and be less volatile.
    \paragraph{Cycles:} Economic and speculative cycles can cause significant fluctuations in activity levels on the blockchain.

\subsubsection{Economic Implications:}

To address the challenges posed by high demand uncertainty, one intuitive solution is to allow the block size to adjust with a base fee ($\mathcal{P}$) and better match the fluctuating demand. When demand spikes, increasing the block size can help accommodate more transactions, alleviating congestion.

\subsection{Uncertainty in Validator Costs}

Validator costs play an important role in the operation of blockchains. These costs can vary significantly depending on the consensus mechanism.

\subsubsection{Definition and Examples}

The marginal cost of transaction validation, denoted by $C'(q)$, is represented by $\eta$. In practice, the covariance between $\Psi$ (user demand uncertainty) and $\eta$ (validator marginal costs) is often positive, implying that the costs to validators also tend to rise as demand increases. This relationship can be mathematically expressed as:
\begin{align}
    Cov[\Psi, \eta] > 0
\end{align}

For example, in PoW blockchains like Bitcoin, the hash rate—a proxy for computational effort and energy consumption—typically increases with higher transaction demand, reflecting a positive and high covariance. Conversely, in PoS systems, the marginal cost of block production is relatively constant and less dependent on fluctuating transaction volumes since the cost to validators is mainly the fixed cost of staking.

\subsubsection{Economic Implications:}

A large positive covariance between validator marginal costs and user demand favors quantity controls $\mathcal{Q}$  In particular, under a price control mechanism ($\mathcal{P}$), validators may face higher costs exactly when demand—and hence their workload—increases. This scenario could lead to inefficiencies where blocks become more expensive to produce precisely when they are most needed.

\subsection{Cryptocurrency Price Fluctuations}

Cryptocurrencies are notoriously volatile, which presents unique challenges for blockchain transaction fee mechanisms. 

\subsubsection{Definition and Examples:}

In economic terms, people generally value their wealth in stable currencies like the dollar or in terms of real goods rather than in the native tokens of blockchains. One measure of price volatility for Ethereum, for instance, is the variance of its token
$Var[\$ETH]$.
This variability means that transaction fees' real (USD) cost can fluctuate widely, even if the fee in native tokens remains constant.

\subsubsection{Economic Implications:}

Large fluctuations in cryptocurrency prices tilt the balance against $\mathcal{P}$ mechanisms and towards $\mathcal{Q}$ mechanisms  During periods when the value of a cryptocurrency like Ethereum is high, the corresponding USD value of base fees in a $\mathcal{P}$ mechanism can become prohibitively expensive. Another interpretation of this result is that if blockchain designers care about implementing a $\mathcal{P}$ mechanism, they should consider allowing fees to be paid in USD and stablecoins.

\subsection{Price Elasticity of Demand}

\subsubsection{Definition and Examples:} The price elasticity of demand for inclusion in the next block refers to the responsiveness of the number of transactions (i.e., block space used) to changes in the base fee. Mathematically, the exponent $\varepsilon$ of our demand function denotes the price elasticity of demand. A high value of $\varepsilon$ suggests that users are highly sensitive to changes in transaction costs. For instance, fast blockchains with short block times or confirmation times, such as Ethereum, exhibit higher price elasticities. The estimate of $\varepsilon_{\textrm{ethereum}} \approx 12.6$ in ~\cite{ndiaye2023bitcoin} suggests that even minor adjustments in the base fee can lead to significant changes in the demand for block space.

\subsubsection{Economic Implications:}

A high price elasticity of demand for inclusion in the next block amplifies gains from $\mathcal{P}$ controls. A higher elasticity facilitates faster adjustments in block size in response to changes in demand, amplifying the benefits of dynamically matching block space to user needs.

\subsection{Decentralization}

\subsubsection{Definition and Examples:}

Decentralization is an aspect of blockchains, often measured by the distribution of power among participants in the network. Validator bargaining power, denoted by $\beta$, inversely correlates with the level of decentralization within the network. A lower value of $\beta$ implies higher decentralization, indicating reduced power concentration among validators.

\subsubsection{Economic Implications:}

In a more decentralized blockchain, the reduced bargaining power of validators makes it easier for blockchain designers to implement and enforce their preferred policies, such as adjustments to block size. This fact is particularly relevant when designing mechanisms to adjust block size in response to fluctuating demand dynamically, ensuring the network can efficiently respond to user needs without undue influence from a concentrated group of validators.

\section{Conclusion}\label{sec: conclusion}

This paper has examined various factors influencing the decision between price-based ($\mathcal{P}$) and quantity-based ($\mathcal{Q}$) transaction fee mechanisms in blockchain systems. Below is a summary table that encapsulates the main findings:

The choice between implementing a policy akin to Ethereum's EIP-1559 ($\mathcal{P}$) or opting for a traditional block size limit ($\mathcal{Q}$) hinges on the blockchain's technical specificities. We found that high demand uncertainty and faster blockchains should favor price controls, while highly decentralized and PoW blockchains should favor simple block size limits. Furthermore, EIP-1559-type mechanisms, when needed, should compute base fees in USD due to cryptocurrency price volatility. The relative balance between five factors -demand uncertainty, validator cost uncertainty, token price volatility, demand elasticity, and decentralization- determines the optimal transaction fee mechanism, which aligns the objectives of welfare-maximizing blockchain designers and profit-maximizing validators.

The optimal choice between a price-based and a quantity-based transaction fee mechanism is not one-size-fits-all but depends on a blockchain’s specific technical characteristics and the economic environment in which it operates. Future research could further quantify these choices and explore optimal supply schedules that incorporate the advantages of both $\mathcal{P}$ and $\mathcal{Q}$ mechanisms, with the potential to offer more flexible and robust fee structures as blockchain technology evolves and matures.

\bibliographystyle{splncs04}
\bibliography{blockchain}

\begin{thebibliography}{1}
\providecommand{\url}[1]{\texttt{#1}}
\providecommand{\urlprefix}{URL }
\providecommand{\doi}[1]{https://doi.org/#1}

\bibitem{akbarpour2020credible}
Akbarpour, M., Li, S.: Credible auctions: A trilemma. Econometrica  \textbf{88}(2),  425--467 (2020)

\bibitem{buterin2018blockchain}
Buterin, V.: Blockchain resource pricing. URL: https://ethresear. ch/uploads/default/original X  \textbf{2} (2018)

\bibitem{chung2023foundations}
Chung, H., Shi, E.: Foundations of transaction fee mechanism design. In: Proceedings of the 2023 Annual ACM-SIAM Symposium on Discrete Algorithms (SODA). pp. 3856--3899. SIAM (2023)

\bibitem{klemperer1989supply}
Klemperer, P.D., Meyer, M.A.: Supply function equilibria in oligopoly under uncertainty. Econometrica: Journal of the Econometric Society pp. 1243--1277 (1989)

\bibitem{ndiaye2023bitcoin}
Ndiaye, A.: Why bitcoin and ethereum differ in transaction fees: A theory of blockchain fee policies. Available at SSRN  (2023)

\bibitem{reis2006inattentive}
Reis, R.: Inattentive producers. The Review of Economic Studies  \textbf{73}(3),  793--821 (2006)

\bibitem{roughgarden2021transaction}
Roughgarden, T.: Transaction fee mechanism design. ACM SIGecom Exchanges  \textbf{19}(1),  52--55 (2021)

\bibitem{weitzman1974prices}
Weitzman, M.L.: Prices vs. quantities. The Review of Economic Studies  \textbf{41}(4),  477--491 (1974)

\end{thebibliography}

\end{document}